# Significantly enhanced critical current densities in MgB$_2$ tapes made by a scaleable, nano-carbon addition route

Yanwei Ma[1] [*], Xianping Zhang[1], G. Nishijima[2], K. Watanabe[2], Xuedong Bai[3]

[1] Applied Superconductivity Lab., Institute of Electrical Engineering, Chinese Academy of Sciences, P. O. 2703, Beijing 100080, China

[2] High Field Laboratory for Superconducting Materials, Institute for Materials Research, Tohoku University, Sendai 980-8577, Japan

[3] Institute of Physics, Chinese Academy of Sciences, Beijing 100080, China

* e-mail: ywma@mail.iee.ac.cn

**Abstract:**

Nanocarbon-doped Fe-sheathed MgB$_2$ tapes with different doping levels were prepared by the *in situ* powder-in-tube method. Compared to the undoped tapes, J$_c$ for all the C-doped samples was enhanced by more than an order of magnitude in magnetic fields above 9 T. At 4.2 K, the transport J$_c$ for the 5 at% doped tapes reached $1.85 \times 10^4$ A/cm$^2$ at 10 T and $2.8 \times 10^3$ A/cm$^2$ at 14 T, respectively. Moreover, the critical temperature for the doped tapes decreased slightly. Transmission electron microscopy showed a number of intra-granular dislocations and the dispersed nanoparticles embedded within MgB$_2$ grains induced by the C doping. The mechanism for the enhancement of flux pinning is also discussed. These results indicate that powder-in-tube-processed MgB$_2$ tape is very promising for high-field applications.

The discovery of superconductivity at 39 K in the $MgB_2$ compound has generated great interest in the field of applied superconductivity[1]. Compared to conventional metallic superconductors, $MgB_2$ has advantages of high transition temperature ($T_c$) and low raw material costs of both B and Mg. In particular, recent studies found a significant upper critical field $H_{c2}$ enhancement, $H_{c2}^{\perp}$ (4.2 K) ≈ 35 T and $H_{c2}^{//}$ (4.2 K) ≈ 51 T, in C-alloyed films. If fitting to data using the two-gap model of $H_{c2}$, $H_{c2}^{//}$ (0) may approach the paramagnetic limit of ~ 70 T[2]. Such critical field properties exceed those of any $N_b$-based conductor at any temperature, suggesting that $MgB_2$ could be a viable replacement for $Nb_3Sn$ as a high field magnet conductor. Indeed, superconducting $MgB_2$ tape has been regarded as one of the most promising materials for the next generation of superconductor applications[3]. The method commonly used to fabricate $MgB_2$ tape is the powder-in-tube technique[4-6].

So far, enormous efforts have been directed towards improving critical current density ($J_c$) through development and application of various novel techniques for fabrication of technically usable materials, such as doping with elements or compounds[7-10], hot isostatic pressing[11], irradiation with heavy ions[12]. Among all the methods, chemical doping is the most promising way for large scale applications. Many dopants have been attempted in order to increase $H_{c2}$ or introduce pinning centers to improve $J_c$, notably in high magnetic fields (H).

Especially, many groups have focused on studying the effect of C-doping on superconductivity in $MgB_2$ compound, since C element is widely recognized to enter the structure through replacing boron[13-16]. Several groups have also reported that the nanoscale SiC-doped $MgB_2$ wires and tapes exhibited much higher $J_c$ values of $MgB_2$ than those of undoped ones[7, 17]. In fact there is increasing suspicion that one of the beneficial effects of SiC addition occurs by C doping of the $MgB_2$[18]. Therefore, from the application point of view, C doping of $MgB_2$ is a quite useful means of alloying $MgB_2$ for enhancing $H_{c2}$ and flux pinning. Actually, carbon[19], diamond[9] and carbon nanotube[20] have recently been added into $MgB_2$, and all of them showed a positive effect in enhancing the $J_c$ property of $MgB_2$ superconductors through magnetic measurement (no transport $J_c$ data is available). However, the effect of pure nano-C doping on $J_c$-H properties of Fe-sheathed $MgB_2$ tapes has not been reported. In this work we will present new results on doping $MgB_2$/Fe tapes with nanometer C particles to achieve an improvement of critical current density by more than one order of magnitude in high magnetic fields.

Powders of magnesium (99.8%, -325 mesh), amorphous boron (99.99%) and carbon nanoparticle powders (20-30 nm, amorphous) were used for the fabrication of tapes by the *in-situ* powder-in-tube method. The C doping ratio was varied from 2.5 to 20 at%. The sheath materials chosen for this experiment were commercially available pure Fe. The mixed powder was filled into a Fe tube of 8 mm outside diameter and 1.5 mm wall thickness in air. After packing, the tube was rotary swaged to two rod types of 5 mm (Batch I) and 3 mm (Batch II) in diameter and then drawn to wires of 1.5 mm in diameter. The wires were subsequently rolled to tapes of ~3.2 × 0.5 mm. Short samples (~ 4 cm each), cut from the tapes, were wrapped in Ta foil and heat treated at 650°C or 750°C for 1 h in flowing high purity Ar, and then cooling in the furnace to room temperature. Undoped tapes were also prepared under the same conditions for use as reference samples.

The phase composition and microstructure were investigated by X-ray diffractometry (XRD) and transmission electron microscopy (TEM). Magnetization measurements were carried out with a superconducting quantum interference device magnetometer. The transport $J_c$ at 4.2 K and its magnetic field dependence were evaluated by a standard four-probe technique with a criterion of 1 μV/cm. The critical current $I_c$ measurement was performed for several samples to check reproducibility.

Figure 1 shows the x-ray diffraction patterns of the series of *in-situ* processed $MgB_2$ tapes with different C doping as well as the XRD pattern of the starting nano-C powder. As can be seen, the undoped samples consist of a main phase, $MgB_2$, with a small amount of MgO present. In the C-doped samples, extra peaks of $Mg_2C_3$ appear as impurity phases, which increase as the doping level increases. Note that there is no peaks related to C in the XRD patterns of the C-doped tapes due to the amorphous C powder used. At the same time, by increasing C doping level, there is a significant shift of the (110) peaks to the higher angle that indicated a distortion of lattice parameter. These results suggest that the substitution of C in the B site actually occurred, which is consistent with recent results of $MgB_2$ bulks doping with C or carbon nanotube[19, 20].

Figure 2 shows the transition curves of tapes with different C doping level determined by susceptibility measurements. The $T_c$ onset for the undoped tapes is ~ 36.5 K. The $T_c$ decreased with increasing nano-C doping level. However, $T_c$ has slightly dropped by 3.5 K for the 15 % high C-doped tapes while only 1.5 K decrease in $T_c$ was observed in the 5% doped samples. This clearly indicates that C doping has little effect on $T_c$, which is in good agreement with recent paper[19]. As reported by Wilke et al.[14], $T_c$ may be used as an indicator of how much carbon is incorporated into the $MgB_2$. Therefore, these results suggest that some amounts of C powders were substituted in the B position in our samples.

In order to characterize the critical current density properties, transport measurements were carried out. Figure 3 shows the transport $J_c$ at 4.2 K in magnetic fields for Fe-sheathed $MgB_2$ tapes with various amounts of nano-C doping from 0 to 20 at% that were heat-treated at 650°C (Batch I). The $J_c$ values of the 5% C doped tape that was sintered at 750°C are also included. Only data above 6 T are shown, because at lower field region, $I_c$ was too high to be measured. The striking result of Fig. 3 is that all the doped tapes showed an enhancement of more than one order of magnitude in $J_c$-H performance in fields above 9 T. The highest $J_c$ value of the Fe-sheathed tapes was achieved in the 5 % nano-C addition, then further increasing C doping ratio caused a reduction of $J_c$ in magnetic fields below 10 T. At 4.2 K and 10 T, $J_c$ for the 5 % doped samples is ~ 21 times larger than that of the undoped tapes. At higher doping levels (x > 5 %), although the $J_c$ in low-field region was depressed, the rate of $J_c$ drop is much slower than for all other samples, clearly indicating strong flux pinning induced by the C doping. The higher the doping lever is, the stronger the flux pinning in high fields. Besides, the $J_c$ values of the 5% C doped sample were much enhanced when the sintering temperature was further increased to 750°C. At the same time, the sample showed slightly better $J_c$ field performance too. High $J_c$ values of $1.1 \times 10^4$ A/cm$^2$ at 10 T and $1.4 \times 10^3$ A/cm$^2$ at 14 T were observed for the samples sintered at 750°C.

Figure 4 shows the magnetic-field dependence of the transport $J_c$ at 4.2 K for pure and 5% C-doped MgB$_2$/Fe tapes annealed at 750°C (Batch II). Several conclusions can be drawn from Fig.4. First, high $J_c$ values were achieved in both the undoped and the C-doped tapes in terms of Batch II samples. Second, again all the $J_c$(H) curves for doped tapes have a much higher $J_c$, more than an order of magnitude larger than for the undoped sample at fields above 10 T. The doped tapes heated at 750°C reveals the highest $J_c$ values compared to all other samples in our experiment: at 4.2 K, the transport $J_c$ reached $1.85 \times 10^4$ A/cm$^2$ at 10 T and $2.8 \times 10^3$ A/cm$^2$ at 14 T, respectively. Even for the C-added tapes sintered at 650°C for 1 h, we observed $J_c$ values of $1.5 \times 10^4$ at 10 T and $1.6 \times 10^3$ A/cm$^2$ at 14 T, respectively. The $J_c$-H properties of our C-doped MgB$_2$ tapes are much better than the MgB$_2$ bulk or wires made by nano-carbon[19] and carbon nanotube[21] doping reported so far, and also are comparable to the best results recently achieved in the nano-SiC doped MgB$_2$ tapes[7, 17]. On the other hand, samples heated at 750°C have a better field performance and higher $J_c$ than samples sintered at 650°C, indicating that the sintering temperature has a significant effect on the $J_c$(H) characteristics for C-doped samples. A higher annealing temperature promotes the C substitution reaction for B[15], thus enhancing flux pinning and improving the high-field $J_c$[20]. It should be noted that this behavior is quite different from that of SiC-doped samples, in which the $J_c$ is insensitive to the sintering temperature above 650°C[21].

When comparing the tapes between Batch I and Batch II, we should note the $J_c$ difference. The $J_c$ values of the first batch tapes were lower than those of the corresponding second batch samples. This $J_c$ difference can be explained by the difference in packing density of MgB$_2$ between the samples. With increasing the swage percentage reduction, the stress applied to the MgB$_2$ during the swaging increases, and a higher packing density of MgB$_2$ is obtained. Actually, from Figs. 3 and 4, the $J_c$-H curves of Batch II tapes were shifted upward compared to Batch I samples, but the field dependence of $J_c$ was not changed, suggesting that the $J_c$ increase for Batch II is clearly attributable to the improvement of the grain connectivity as a consequence of densification of the tape core.

In order to understand the mechanisms for the enhancement of $J_c$ at higher fields in the nano-C-doped samples, a TEM study was performed. Figure 5 shows the typical low magnification and high-resolution TEM micrographs for the C-doped tapes. TEM images showed that the grains size is less than 100 nm, but in some areas this is hard to distinguish as the grains are so well consolidated. This is because partial melting occurs in the C-doped sample, resulting in an excellent grain connection. Both the selected diffraction (SAD) patterns, shown in the corner of the Fig.5a and b, consist of well defined ring patterns, meaning a very fine grain size. However, the SAD patterns also suggest that the samples sintered at 750°C were very well crystallized compared to tapes heated at 650°C. More notably, the TEM examination revealed that there are a number of impurity phases in the form of nanometer-size inclusions (1-20 nm in size) inside grains in the nano-C-doped samples (see Fig.5a and b). The higher the sintering temperature is, the more the finer nanoscale particles scattering within the grain (Fig.5b). Some of nanoparticles embedded in the $MgB_2$ grains have a thin and clear interface boundary (Fig.5c). The energy dispersive spectroscopy analysis of the grains revealed the presence of uniformly distributed Mg, B, C and O (Fig.5d). This suggests that the inclusion nanoparticles might be unreacted C, $BO_x$, and BC or MgO and $Mg_2C_3$ detected by XRD. In addition, a large number of dislocations are present in the $MgB_2$. Some of these dislocations are indicated by white arrows. The high defect density is consistent with the reduced $T_c$s of the C-doped samples. These nanosized inclusions and intragrain defects created by C doping can serve as strong pinning centers to improve flux pinning. This is clearly demonstrated by the superior $J_c$–H performance of the C-doped samples, as shown in Figs. 3 and 4. Accordingly, a high density of flux-pinning centers and good grain connection for the $MgB_2$ phase are responsible for the excellent performance in our doped tapes.

In summary, by utilizing inexpensive nano-C doping, $J_c$ enhancements of up to over an order of magnitude in high field region were achieved in a reproducible way for Fe-sheathed $MgB_2$ tapes made by the *in situ* powder-in-tube method. This significantly improves the potential of $MgB_2$ for many applications. It is expected that further improvement in $J_c$-H is expected by optimizing the microstructure and the sintering temperature.

**Acknowledgments**

We thank Yulei Jiao, Ling Xiao, Xiaohang Li, Hongli Suo, Haihu Wen and Liye Xiao for their help and useful discussions. This work is partially supported by the National Science Foundation of China (NSFC) under Grant No.50472063 and No.50377040 and National "973" Program (Grant No. 2006CB601004).

# Captions

Figure 1 XRD patterns of undoped and nano-C-doped tapes. The data were obtained after peeling off the Fe-sheath. The peaks of Fe were contributed from the Fe sheath.

Figure 2 Normalized magnetic susceptibility versus temperature for the samples with different doping level.

Figure 3 Transport critical current densities of Fe-sheathed tapes with nano-C doping level from 0 to 20 at%, which were heat-treated at 650°C (Batch I), at 4.2 K in magnetic fields. The $J_c$ values of the 5% C doped tape that was sintered at 750°C are also included.

Figure 4 Transport $J_c$-H properties at 4.2 K for Fe-sheathed undoped and 5 at% nanosized C-doped tapes (Batch II).

Figure 5 TEM micrographs showing the nanoparticle inclusions and dislocations of the C-doped samples. Some of the dislocations have been labelled using white arrows. The diffraction pattern of each sample is shown in the corner of the corresponding image. a, Low-magnification micrograph for samples sintered at 650°C. b, High-magnification micrograph for samples sintered at 750°C. c, Image of nanoparticle showing lattice fringes. d, the energy dispersive spectroscopy element analysis of $MgB_2$ grains.

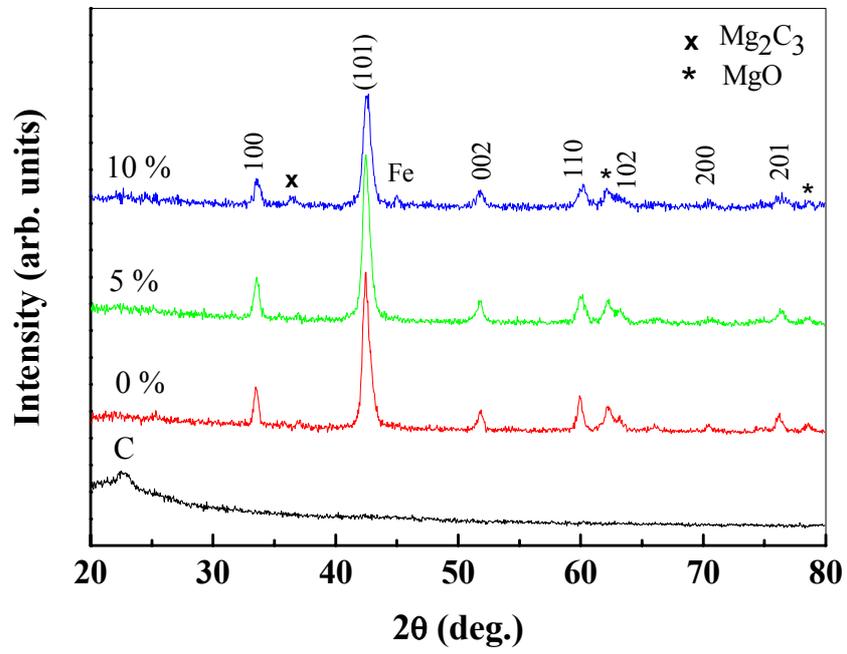

Fig.1 Ma et al.

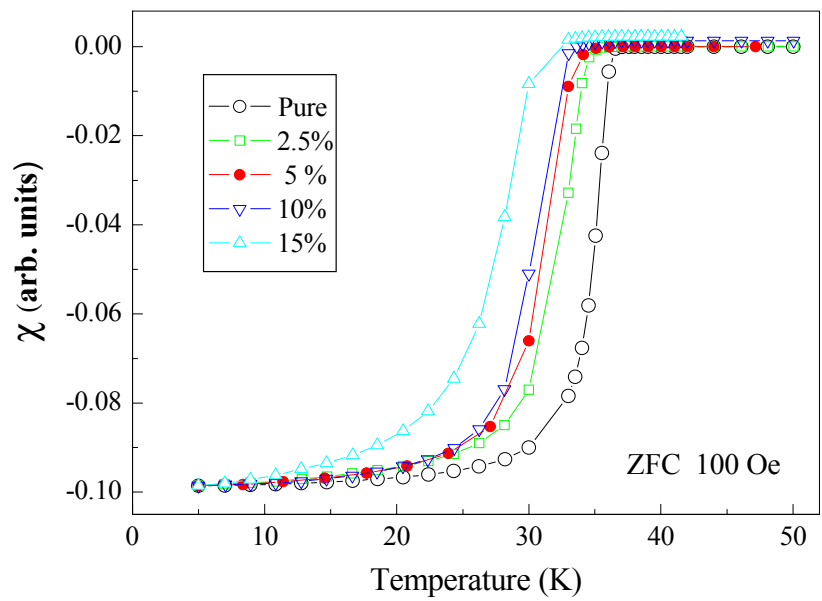

Fig.2 Ma et al.

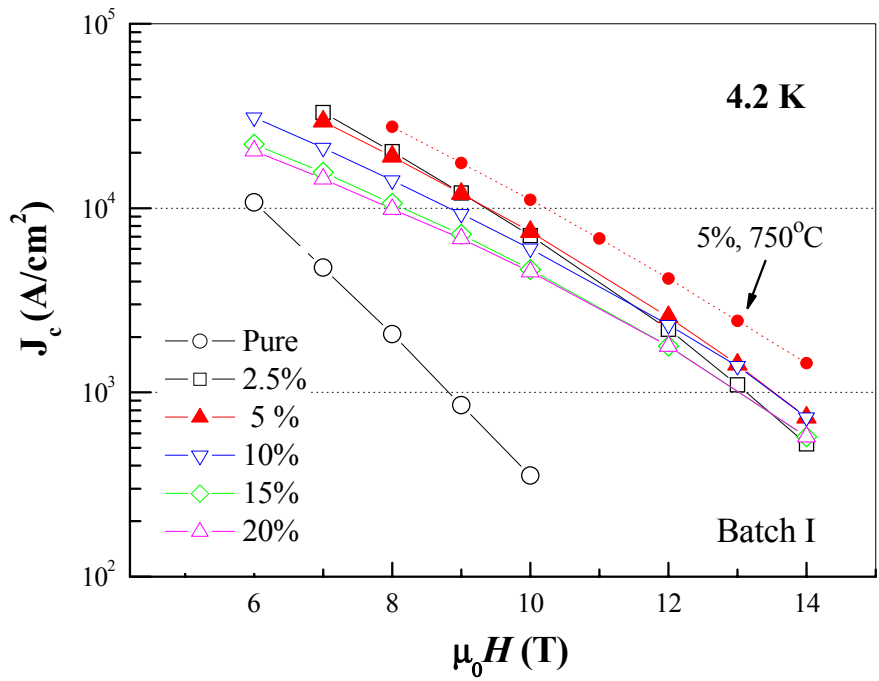

Fig.3 Ma et al.

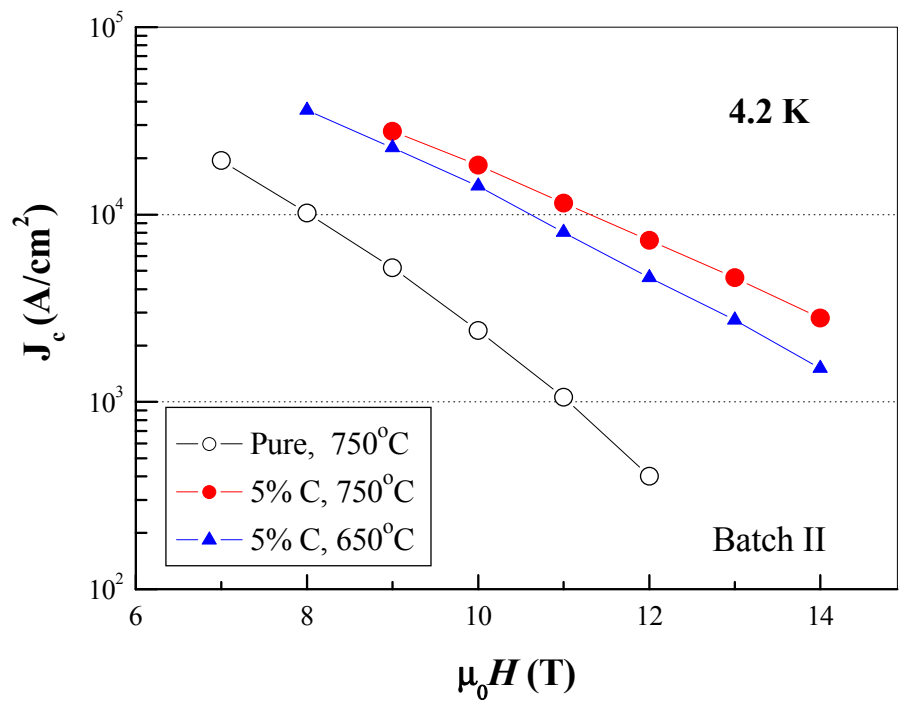

Fig.4 Ma et al.

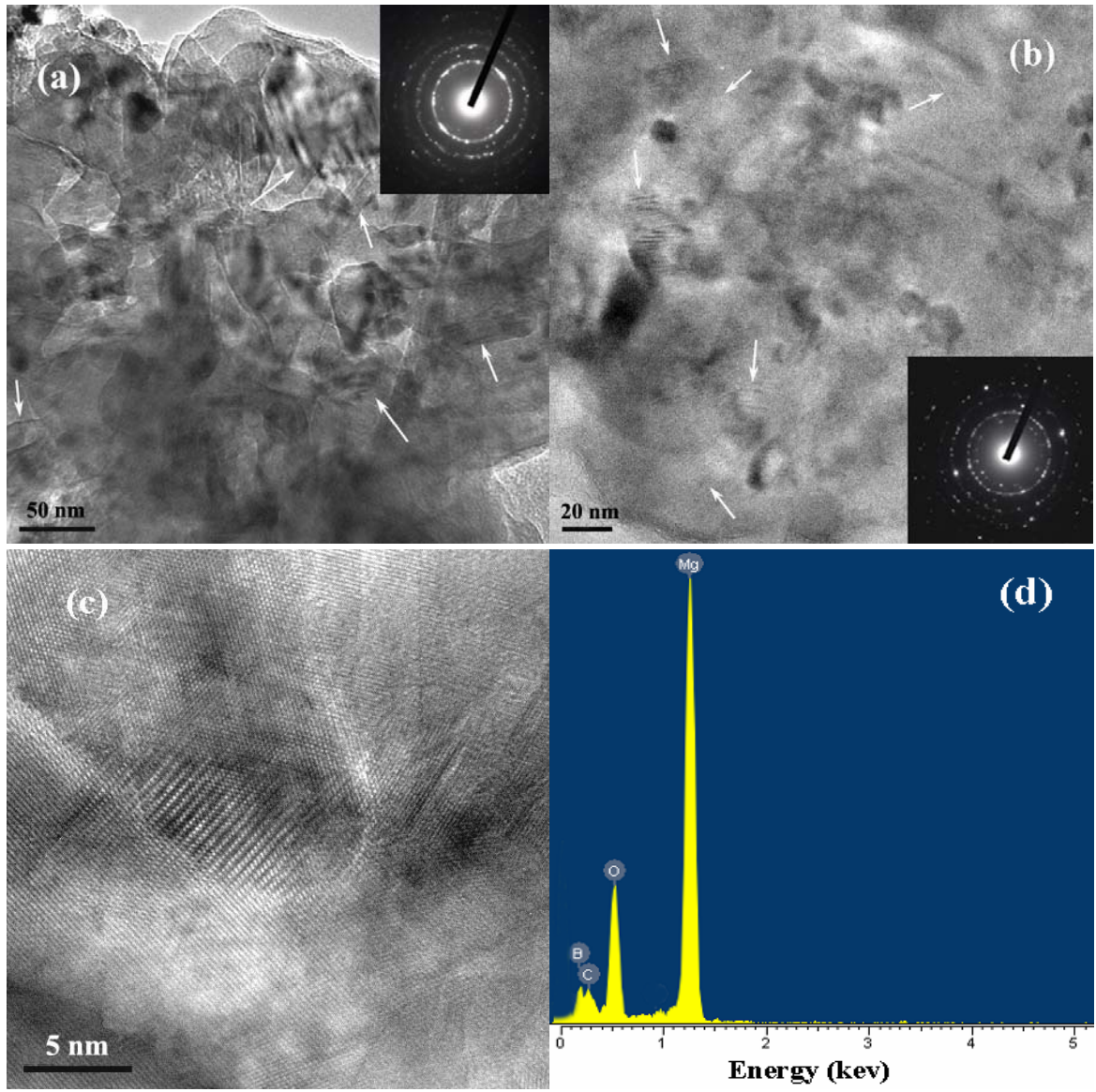

Fig.5 Ma et al.